\documentclass[twocolumn]{aastex631}
\usepackage{amsmath}
\usepackage{hyperref}
\usepackage{soul}
\usepackage{multirow}

\begin{document}

\title{Detection of very high-energy gamma-ray emission from the radio galaxy M87 with LHAASO}


\author{Zhen Cao}
\affiliation{Key Laboratory of Particle Astrophysics \& Experimental Physics Division \& Computing Center, Institute of High Energy Physics, Chinese Academy of Sciences, 100049 Beijing, China}
\affiliation{University of Chinese Academy of Sciences, 100049 Beijing, China}
\affiliation{TIANFU Cosmic Ray Research Center, Chengdu, Sichuan,  China}
 
\author{F. Aharonian}
\affiliation{Dublin Institute for Advanced Studies, 31 Fitzwilliam Place, 2 Dublin, Ireland }
\affiliation{Max-Planck-Institut for Nuclear Physics, P.O. Box 103980, 69029  Heidelberg, Germany}
 
\author{Axikegu}
\affiliation{School of Physical Science and Technology \&  School of Information Science and Technology, Southwest Jiaotong University, 610031 Chengdu, Sichuan, China}
 
\author{Y.X. Bai}
\affiliation{Key Laboratory of Particle Astrophysics \& Experimental Physics Division \& Computing Center, Institute of High Energy Physics, Chinese Academy of Sciences, 100049 Beijing, China}
\affiliation{TIANFU Cosmic Ray Research Center, Chengdu, Sichuan,  China}
 
\author{Y.W. Bao}
\affiliation{School of Astronomy and Space Science, Nanjing University, 210023 Nanjing, Jiangsu, China}
 
\author{D. Bastieri}
\affiliation{Center for Astrophysics, Guangzhou University, 510006 Guangzhou, Guangdong, China}
 
\author{X.J. Bi}
\affiliation{Key Laboratory of Particle Astrophysics \& Experimental Physics Division \& Computing Center, Institute of High Energy Physics, Chinese Academy of Sciences, 100049 Beijing, China}
\affiliation{University of Chinese Academy of Sciences, 100049 Beijing, China}
\affiliation{TIANFU Cosmic Ray Research Center, Chengdu, Sichuan,  China}
 
\author{Y.J. Bi}
\affiliation{Key Laboratory of Particle Astrophysics \& Experimental Physics Division \& Computing Center, Institute of High Energy Physics, Chinese Academy of Sciences, 100049 Beijing, China}
\affiliation{TIANFU Cosmic Ray Research Center, Chengdu, Sichuan,  China}
 
\author{W. Bian}
\affiliation{Tsung-Dao Lee Institute \& School of Physics and Astronomy, Shanghai Jiao Tong University, 200240 Shanghai, China}
 
\author{A.V. Bukevich}
\affiliation{Institute for Nuclear Research of Russian Academy of Sciences, 117312 Moscow, Russia}
 
% ...


\begin{abstract}
The nearby radio galaxy M87 is a very-high-energy (VHE) gamma-ray emitter established by observations with ground-based  gamma-ray detectors.
Here we report the long-term monitoring of M87 from 2021 to 2024 with  Large High Altitude Air Shower Observatory (LHAASO).  M87 has been detected by LHAASO with a statistical significance   $\sim 9\sigma$. The observed energy spectrum extends to 20 TeV, with a possible hardening at $\sim 20$ TeV and then a clear softening  at  higher energies. Assuming that the intrinsic spectrum is described by a single power law up to 20 TeV,  a tight upper bound on the extragalactic background light (EBL) intensity is obtained.
A strong VHE flare lasting eight days, with the rise time of $\tau_{r}^{\rm rise} = 1.05\pm0.49$~days  and  decay time of $\tau_{d}^{\rm decay} = 2.17\pm0.58$~days, was found in early 2022. A possible  GeV flare is  seen also in the Fermi-LAT data during the VHE flare period.
The variability time as short as one day seen in the LHAASO data suggests a compact emission region with a size of  $\sim 3\times 10^{15}\delta\, {\rm cm}$ ($\delta$ being the Doppler factor of the emitting region), corresponding to  a few Schwarzschild radii of the central supermassive black hole in M87. 
The continuous monitoring  of the source reveals a duty cycle of $\sim 1\%$ for VHE flares with a flux above $ 10^{-11}{\rm~erg~cm^{-2}~s^{-1}}$.

\end{abstract}

\keywords{Active galactic nuclei,  High energy astrophysics}

\section{Introduction}           
\label{sect:intro}

The radio galaxy M87 is one of the nearest active galactic nuclei (AGN) located in the Virgo Cluster at a distance of approximately $16.8\pm0.8$ Mpc \citep{2009ApJ...694..556B}. It hosts a super massive black hole (SMBH),  known as M87$^\star$,  with a  mass of $(\rm 6.5\pm0.2_{stat}\pm0.7_{sys}) \times 10^{9}~M_{\odot}$ \citep{2019ApJ...875L...6E}. M87 has been a subject of extensive studies across
the entire electromagnetic spectrum (see \citet{2021ApJ...911L..11E} for a review of recent observations).

The first  evidence for very-high-energy (VHE) gamma-ray emission from M87 was reported by the High Energy Gamma Ray Astronomy
during 1998-1999 \citep{2003A&A...403L...1A}. It was later confirmed by the next-generation imaging air Cherenkov
telescopes (IACTs), including H.E.S.S. \citep{2006Sci...314.1424A}, VERITAS \citep{2009Sci...325..444A,2010ApJ...716..819A}, MAGIC \citep{2008ApJ...685L..23A}, as well as by the High Altitude Water Cherenkov Observatory (HAWC)  at a marginal statistical level \citep{2021ApJ...907...67A}. M87 has also been detected by Fermi-LAT in the high-energy (HE, $0.1-30{~\rm GeV}$) band \citep{2009ApJ...707...55A}.

The exact location of the VHE gamma-ray emitting region in M87 remains unknown. The angular resolution of ground-based VHE instruments is of the order of 0.1 degree (corresponding to 30 kpc projected size) and, therefore, does not allow to locate the emitting region. The suggested sites of TeV gamma-ray production range from  the immediate environment (e.g., the magnetosphere) of the supermassive black hole\citep{2006Sci...314.1424A,2007ApJ...671...85N,2012MPLA...2730030R} to inner (subparsec) parts of
the jet \citep{2004A&A...419...89R,2005ApJ...634L..33G},  a compact hot spot (the so-called HST-1 knot) at a distance of 100~pc
along the jet \citep{2006MNRAS.370..981S}, and large-scale
structures in the kiloparsec jet \citep{2005ApJ...626..120S}. To further investigate the location of the VHE gamma-ray emission site, variability studies and the search for correlations with other wavelengths have usually been used  \citep[e.g.,][]{2009Sci...325..444A}.

Four major VHE flares from M87 have been reported before the current observations, in 2005 \citep{2006Sci...314.1424A}, 2008 \citep{2008ApJ...685L..23A,2009Sci...325..444A}, 2010 \citep{2012ApJ...746..151A} and 2018 \citep{2024A&A...685A..96H,2024arXiv240417623T}. The variability time scales of the these VHE flares is about one day. This points,
through the causality argument, to a compact emission
region with a size smaller than  $3\times 10^{15} \delta \, {\rm cm}$, where $\delta\sim  {\rm a \,\, few }$ is the Doppler factor of the emitting region.
The coincidence of the first VHE flare with a giant X-ray flare
from HST-1 \citep{2006ApJ...640..211H}  initially
prompted speculations that the observed VHE emission may have originated in
HST-1  (e.g., \citet{2006MNRAS.370..981S}).  
During the second flaring episode, however, HST-1 was in a low flux state, while  radio measurements showed a flux increase in the core region within a few hundred Schwarzschild radii of the SMBH, suggesting
the direct vicinity of the SMBH as the site of the VHE gamma-ray emission \citep{2009Sci...325..444A}. This conclusion was further supported by the detection of an enhanced X-ray flux from the core  by Chandra.
Therefore, the potential site of such rapid flares is related to either the core
of M87 \citep{2006Sci...314.1424A} or to very compact regions in the jet, for instance, the HST-1 knot \citep{2006ApJ...640..211H}. Note that, however,  the location of HST-1, at a distance of 100 pc from the central
BH, would require an extremely tight collimation of the jet to account for the observed
fast (days-scale) TeV variability \citep{2006Sci...314.1424A}.  Moreover, the fact that the Chandra X-ray flux from HST-1 did not change
much during  subsequent VHE flaring episodes (after the one in 2005)  has  been taken to disfavor HST-1 as a site of the rapid TeV flaring activity \citep{2012MPLA...2730030R}.

With a large detector area and excellent gamma-ray/background discrimination power, the sensitivity of  Large High Altitude Air
Shower Observatory (LHAASO) at VHE are  higher than any other extensive air shower (EAS) experiments.
Moreover,  in contrast to Atmospheric Cherenkov telescopes,  LHAASO can monitor M87 continuously, thanks to {its} wide field of view and high duty cycle. 
Here we present the three-year continuous observations  of M87 from 2021 to 2024 with LHAASO. During the campaign, a  major VHE flare was detected in 2022. Characteristics of the VHE flare are investigated and possible correlation with the {GeV} band measured by Fermi-LAT is discussed. This paper is organized as follows. In Section 2, we present the LHAASO observations  of M87 and the analysis methods. 
In Section 3, the results of Fermi-LAT during the same observation period are presented. 
In Section 4, we discuss the implications of our findings, and summarize the results in Section 5.

\section{LHAASO observations}
\label{sect:lhaaso}

LHAASO is a multifaceted and comprehensive EAS detector array. The facility consists of three  sub-arrays, including the Water Cherenkov Detector Array (WCDA, sensitive to VHE gamma-rays at energy from a few hundreds of GeV to tens of TeV), the 1.3 km$^2$ ground array (KM2A, sensitive to VHE gamma-rays  at energy above 20~TeV) and the Wide Field-of-view Air Cherenkov/Fluorescence Telescope Array (WFCTA), designed for cosmic ray studies from 10~TeV to 1~EeV. The detailed information about the performance of these sub-arrays and data reconstruction algorithms can be seen in \citet{2019arXiv190502773C,2024ApJS..271...25C}.
The details of data analysis are summarized in Appendix \ref{sect:app_haaso}. The first source catalog of Large High Altitude Air Shower Observatory reported the detection of a very-high-energy gamma ray source, 1LHAASO J1219+2915 \citep{2024ApJS..271...25C}, which is identified as the counterpart of  a low-luminosity AGN \citep[NGC~4278][]{Cao_2024b}.

The data utilized for M87 in this work is collected by full array of LHAASO-WCDA from 8 March, 2021 to 16 March 2024 and full array of LHAASO-KM2A from 20 July, 2021 to 31 January, 2024. The effective live time is 1026\,days and 884\,days, respectively. The number of fired hit ($N_{\rm{hit}}$) is selected as the estimator of primary energy for WCDA \citep{Aharonian_2021a}. The events are binned into seven segments with $N_{\rm{hit}}$ value of [60, 100), [100, 200), [200, 300), [300, 500), [500, 700), [700, 1000) and [1000, 2000]. The KM2A events are divided into bins with reconstructed energy width of 0.2 in log scale \citep{Aharonian_2021b}. The ``direct integral method"  is adopted here to estimated cosmic ray background \citep{Fleysher_2004}. 
A binned maximum likelihood analysis based on the forward folding method ({ i.e., convolving the assumed spectrum model with the detector response})  is performed to estimate best-fit values of position and spectrum parameters. The test statistic (TS) is used to calculate significance of the target source, $\rm{TS} = -2(\ln \mathcal{L}_{\rm{b}} - \ln \mathcal{L}_{\rm{s+b}})$, where $\mathcal{L}_{\rm{b}}$ and $\mathcal{L}_{\rm{s+b}}$  are the likelihood values for the background without (null hypothesis) and with the target signal model, respectively.

The {LHAASO-WCDA} full-time significance map of the M87 region is shown in the left panel of Figure \ref{SigMap_WCDA}. 
The gamma-ray spatial distribution is well described by a point-like source model with a TS of 87.4 (corresponding to $\sim 9\,\sigma$). The $\Delta \rm{TS}$ value between the point-like model and 2-D Gaussian model is smaller than 0.1.
The best-fit position of the source is (R.A. = $187.73^{\circ}\pm0.03^{\circ}$, Decl. = $12.44^{\circ}\pm0.03^{\circ}$). This position is only 0.03$^{\circ}$ away from the radio position of the core of M87, suggesting that the source is spatially associated with M87. 
\subsection{The full-time average spectrum}
{We combined the WCDA and KM2A data using a joint forward folded fit to determine the observed spectral energy distribution (SED) of
M87,  shown in Figure \ref{SED_WCDA_Obs}, assuming  a power-law with exponential cutoff (denoted
as PLEC) function or a log-parabola (LP) function
(denoted as LP). 
The PLEC function is defined as
\begin{equation}
	\label{eq1}
	{dN(E)}/{dE}=N_0({E}/{\rm 3TeV})^{-\gamma}{\rm exp}(-E/{E_{\rm cut}}),
\end{equation}
and the LP function is defined as 
\begin{equation}
	\label{eq2}
	{dN(E)}/{dE}=N_0({E}/{\rm 3TeV})^{-\gamma-\beta {\rm ln} (E/{\rm 3TeV})},
\end{equation}
where $E_{\rm cut}$ is the cutoff energy, $\gamma$ is the photon index of the spectrum and $\beta$ is  the coefficient reflecting the spectral curvature.
WCDA covers the energy range from 1 to  20
TeV, and KM2A has one detection at about 20 TeV with the upper limits extending to about 50 TeV.
The two measurements are consistent with
each other at 20 TeV. Fitting the
LP function, the yielded $\chi^2/{\rm ndf}$ is 5.7/3, where ${\rm ndf}$ is the number of degrees
of freedom. If fitting with the PLEC function,  the yielded $\chi^2/{\rm ndf}$  is {4.2/3}.

{The cutoff shape  at the high energy in the observed SED could be partly or fully caused by the  extra-galactic background light (EBL) absorption. The intrinsic spectrum can be obtained by correcting for the absorption due to EBL.  The simplest assumption for the intrinsic spectrum is that it is described by a power-law function  $dN/dE=N_{0}(E/{\rm 3TeV})^{-\gamma}$.}
To incorporate the  EBL absorption into our spectral models, we select three models that represent the uncertain range
of the EBL intensity,  the \textit{kneiske} model \citep{2010A&A...515A..19K} for the lower boundary of the EBL flux, 
the \textit{franceschini} model for the medium intensity \citep{2008A&A...487..837F}, and \textit{dominguez-upper} model reflecting upper bounds of the EBL flux \citep{2011MNRAS.410.2556D}. 
{As demonstrated in Table \ref{tab1:EBL_SED}, these PLxEBL models all yield good fit to the observed spectral data with {  slightly improved}  $\chi^2/{\rm ndf}$ compared with the PLEC and LP fit. A {higher} EBL flux results in a harder intrinsic spectrum, although the {statistical} error in the photon index is still large. } For subsequent analysis, we adopt the medium-intensity EBL model (the \textit{franceschini} model)  as the baseline EBL model.
The photon attenuation derived from the EBL model as a function of energy at the distance of M87 (z = 0.0042) is applied to correct the detection efficiency. With EBL-corrected detection efficiency, the photon index of the intrinsic spectrum is 2.37$\pm$0.14 for the full-time period.
The best-fit parameters for the power-law intrinsic spectrum are summarized in Table \ref{tab1:OverallSED}. As shown in the left panel of Figure \ref{SED_WCDA}, the intrinsic spectrum for the full time data  extends to  tens of TeV without a visible softening. Interestingly, a possible spectral hardening is seen at 20 TeV.

{It is possible that the intrinsic spectrum is not a simple power law, but has a steepening at the high-energy end. We assume that  the {\em intrinsic} spectrum is described by a power law with an exponential cutoff or a log-parabola function and test the goodness of the fitting. We find that the results do not improve compared to the simple power law model and the fitting parameters ($\beta$ or $E_{\rm cut}$) are not well constrained (the details are shown in Table~\ref{tab4:SED_model} in the Appendix), which could be due to the low statistics of the data. }

\subsection{The light curve}
To ensure that the TS value of each bin is greater than 4, we show the light curve of M87 during the three-year continuous observations with LHAASO-WCDA in Figure \ref{LC1} (the upper-left panel). 
As the reported variability time scale of M87 could be as short as one day, we also extract the  light curve of M87 binned on  two-days and one-day  using three years of LHAASO-WCDA data, assuming a spectral index equal to the value obtained in the full time analysis.

For the two-day binning light curve, the likelihood variability test \citep{Abeysekara_2017} of a constant flux (set to the average flux of full time) gives a p-value of $4\times10^{-3}$  ($\sim2.7\sigma$), indicating the possible existence of variability. In order to quantify the fluctuations of temporal profiles, we employ the Bayesian block method \citep{2013ApJ...764..167S} on the two-day binning light curve with a prior false positive probability of 5\%. Only one clear block from MJD 59607 to MJD 59615 is found, which is shown in the upper-right panel of Figure \ref{LC1} (the blue line). We identify this period as a flare, whose details will be discussed in section \ref{sect:2022flare}. {Four time window widths of 1 day, 2 days, 4 days, and 8 days were used to search the 1026-day LHAASO-WCDA data, resulting in 2071 searches. A conservative way is adopted to calculate the post-trial significance of the flare by setting the trial number equal to 2071, which yields a post-trial significance of $\sim 4.4\,\sigma$.}

\subsubsection{The 2022 VHE flare}
\label{sect:2022flare}



The best-fit position of the emission detected during the flare period is (R.A. =$187.75^{\circ}\pm0.06^{\circ}$, Decl. = $12.39^{\circ}\pm0.06^{\circ}$), which is 0.07$^{\circ}$ away from radio position of core of M87, as shown in the middle panel of Figure \ref{SigMap_WCDA}. The TS value and photon index of the intrinsic spectrum  for this flare  are 40.1 and $2.57\pm0.23$ (see Table \ref{tab1:OverallSED}), respectively.
The spectrum of the flare is also shown in Figure \ref{SED_WCDA} (the middle panel).  

The VHE flare detected by LHAASO  has an average flux of $(3.78\pm0.71)\times10^{-12}~\rm photons~cm^{-2}~s^{-1}$, corresponding to an energy flux of $1.38\times 10^{-11}{\rm~erg~cm^{-2}~s^{-1}}$ in 1-20 TeV assuming  {a photon index $\gamma=2.57\pm0.23$}. With a distance of $d=16.8\,\mathrm{Mpc}$, the corresponding isotropic VHE luminosity is $\sim4.7\times10^{41}~{\rm erg~s^{-1}}$. This is a non-negligible amount compared to the bolometric  luminosity ($\sim 10^{42}\,\mathrm{erg\,s^{-1}}$) of the M87 nucleus  \citep{2004A&A...425..825K}  and the total kinetic luminosity ($\sim 10^{44} {\rm erg~s^{-1}}$)  of the jet  \citep{2000ApJ...543..611O}.

To derive the variability timescale of M87, the one-day time-bin light curve of the flare is fitted with a two-sided exponential function,
\begin{small}
\begin{equation}
    F = F_{0}\times e^{-|t-t_{0}|/\Delta\tau}{\rm with}
    \left\{
    \begin{aligned} 
        \Delta\tau & = \Delta\tau^{\rm rise} & \textrm{for}\enspace t<t_{0} \\
        \Delta\tau & = \Delta\tau^{\rm decay} & \textrm{for}\enspace t\geq t_{0} .
    \end{aligned}
    \right.
\end{equation}
\end{small}
{The peak time $t_0$ is set to be free, and the  resulting flux doubling time $\tau_{\rm d} = {\rm ln(2)}\times\Delta\tau$} of the rising part is $\tau_{\rm d}^{\rm rise} = 1.05\pm0.49$ days and that of the decaying part is $\tau_{\rm d}^{\rm decay} = 2.17\pm0.58$ days. The variability time scale is similar to that of the 2010 flare \citep{2012ApJ...746..151A}, but  the 2022 flare reported in this work rises faster and decays slower. The fitting result is shown in Figure \ref{LC2}. 
This obtained variability time scale $\tau_{d} = \tau_{d}^{\rm rise} = 1.05\pm0.49 $ days corresponds to the size of a spherical emission region of $R  \leq c\tau_{d}\delta \simeq 2.7 \times10^{15}\delta~\rm cm$, where $\delta$ is the Doppler factor to describe the relativistic beaming.

\subsubsection{The low-state emission}

We consider the non-flare period (i.e., excluding the flare period from MJD 59607 to MJD 59615)  as the low-state. The significance map of this low-state emission is shown in the right panel of Figure \ref{SigMap_WCDA} and the best-fit position, marked by the cyan open cross, is consistent with the radio position of the core of M87.
The TS value of the low-state emission is 76.0. The photon index of the low-state emission, 2.31$\pm$0.17,  is similar to the one obtained with fitting the entire data set.
This photon index is consistent with  that of the low state  emission of M87 measured by H.E.S.S. \citep{2023A&A...675A.138H}. The spectral data for the low-state emission is shown in Figure \ref{SED_WCDA} (the right panel). 


\section{Fermi-LAT observations}
\label{sect:lat}

We searched for multi-wavelength (radio to GeV) data of M87 during the 2022 VHE flare period, and found that only Fermi Large Area Telescope (LAT) had the simultaneous observations in this flare period.

The Fermi-LAT data \citep{2009ApJ...697.1071A} of M87 are extracted from Fermi Science Support Center\footnote{https://fermi.gsfc.nasa.gov/ssc/data/access/}.
The data cover about 15.5 years from August 4, 2008 to March 16, 2024, in an energy range between 100 MeV and 100 GeV. The details of analysis are summarized in Appendix \ref{sect:app_lat}. For the  15.5-year observation of M87, the source is detected with TS=2434.31 (corresponding to a significance of $\sim49\sigma$). The average flux is $F=(1.79\pm0.10)\times 10^{-8}~\rm photons~cm^{-2}~s^{-1}$, 
with a photon index of $\Gamma=2.05\pm0.03$. The average flux is shown by the gray lines in the bottom panels of Figure \ref{LC1}.

We also analyzed three-year data of M87 with the same observation period as LHAASO, i.e., from 8 March, 2021  to 16 March 2024 (MJD 59281--60385). Firstly, we use three-month time bin (the same time bin as used for LHAASO light curve) to obtain the light curves, and the results are shown in the bottom-left panel of Figure \ref{LC1}.
Compared to the average flux, there are no significant variations in the light curve, suggesting no significant evidence of variability on timescales of several months over the three-year period. 
Considering that the VHE flare measured by LHAASO lasts eight days, we also use 8-day time bin (centred on the VHE flare) to obtain the GeV light curves,  shown in the bottom-right panel of Figure \ref{LC1}. 
Interestingly, we find that two time bins have TS values greater than 9, i.e., one before the VHE flare period with TS=10.7 ($\sim3.3\sigma$ significance) and the other one during the same period as the VHE flare with TS=9.5 ($\sim3.1\sigma$ significance).
The GeV flux during the VHE flare period is $(6.80\pm3.26)\times 10^{-8}~\rm photons~cm^{-2}~s^{-1}$, a factor of 3.8 higher than the average flux. 
Given the weakness of the GeV source, shorter-term variations are difficult to probe.
The Fermi-LAT results are summarized in Table \ref{tab2:latspec}.

We compare the GeV spectrum with the VHE spectrum during the flare period and the low-state period in Figure \ref{SED_0}. For the low-state period, the VHE emission is consistent with the extrapolation of the GeV spectrum, indicating possibly a single spectral component for both GeV and VHE emissions. For the flare period, it is possible that the GeV emission can smoothly connect to the VHE emission. However,  since the error of the GeV flux during the flare period is large, one cannot reach a reliable conclusion.



\section{Discussions}

\subsection{VHE flare emission}

For the variability time of $\tau  = 1.05$~days of the current VHE flare,  the size of the emission region should be smaller than $2.7\times 10^{15} \delta \, {\rm cm}$, where $\delta$ is the Doppler factor of the emitting region. The jet of M87 is observed at a large
angle $\theta\sim 20$ degrees \citep{1999ApJ...520..621B}, making M87 a “misaligned” BL Lacertae object  that most  jet emission  is characterized by rather weak Doppler
factors $\delta = 1/[\Gamma_j (1-\beta_j \cos 
\theta)]\la 3$, where $\Gamma_j\sim 10 $ is the jet Lorentz factor.  Since
the mass of the SMBH in M87 is well established, $M=6\times 10^9 M_\odot$, the Schwarzschild radius is estimated quite accurately: $R_s=2GM/c^2\simeq 2\times 10^{15}{\rm cm}$. Therefore, the size of the emission region of the VHE flare is only a few  Schwarzschild radii of the SMBH.

The small size of the TeV emission region, together with the  correlated variability between the X-ray emission (and radio emission) of the core and the TeV emission in some previous VHE flares, suggest that the relevant non-thermal particle acceleration and emission processes could take place in the  core of M87.
One scenario involves inner sub-pc jets that produce VHE emission like blazars \citep{2005ApJ...634L..33G,2008MNRAS.385L..98T}. \citet{2005ApJ...634L..33G} considered the case of a relativistic jet decelerating substantially on sub-parsec distances from the core, and showed that the velocity difference between a faster and a slower portion of the outflow could lead to the enhanced inverse-Compton emission of the former one in the TeV range.

Magnetospheric particle acceleration and emission models have also been invoked to explain the observed VHE emission in M87 \citep{2007ApJ...671...85N,2008A&A...479L...5R}.  Such a mechanism of gamma-ray emission from the vicinity of a BH is a close analog of the mechanism of pulsed gamma-ray emission from the vicinity of neutron stars in pulsars.  Usually, efficient particle acceleration in these
scenarios is related either to gap-type \citep{2007ApJ...671...85N,2011ApJ...730..123L} or centrifugal-type processes occurring in the magnetosphere around a rotating black hole \citep{2008A&A...479L...5R}. Because of its low  infrared luminosity, the nucleus of M87 can be effectively transparent for gamma-rays up to an energy of $\sim 10$ TeV.

Other types of models have been also put forward to explain the observed rapid variability of the VHE $\gamma$-ray emission from M87. Recent high-resolution general-relativistic magnetohydrodynamic simulations show the occurrence of episodic magnetic reconnection events that can power flares near the black hole event horizon \citep{2020ApJ...900..100R,2023ApJ...943L..29H}. It was also argued that interactions of a relativistic  magnetized outflow around the jet formation zone with a star partially tidally disrupted by the interaction with
the SMBH can lead to an efficient production of VHE photons via hadronic processes \citep{2010ApJ...724.1517B,2017ApJ...841...61A}. 


It is worth to mention that that, unlike IACTs, LHAASO data provide 
continuous and uniform TeV measurements of M87. While  IACTs observations  have a better sensitivity for short-time exposure,  
LHAASO observations provide a comprehensive study of the TeV light curve on the months and years time scales.
The measurements presented in this work can  be used to estimate the duty cycle
of VHE flares in M87. {The duty cycle of astrophysical sources is  defined as the fraction of time during which the sources are  in a flaring state. This can be used to estimate the active time of the central engine and the total energy release during the flaring state, in comparison with that in the low-state.   Previous IACT  observations of M87 found that the  respective duty cycle of VHE flares  are $\sim 14\%$, $7\%$ and $4\%$ for a threshold flux of $0.5\times 10^{-11} {\rm  cm^{-2} s^{-1}}$, $0.8\times 10^{-11} {\rm  cm^{-2} s^{-1}}$ and $ 10^{-11} {\rm  cm^{-2} s^{-1}}$\citep{2012ApJ...746..151A}. However, due to the observing strategy  for
M87, i.e., observations have been intensified after the detection
of a high state, the data set could be biased. The derived duty cycle
is, therefore, likely overestimated and should be considered an
upper limit\citep{2012ApJ...746..151A}.  During the three-year continuous monitoring of M87,  LHAASO detected one strong flare over a period of 8 days with an energy flux reaching $1.38\times 10^{-11}{\rm~erg~cm^{-2}~s^{-1}}$ in 1--20~TeV, revealing a duty cycle of $\sim {\rm 8 d/ 3 yr}\sim 1\%$ for such bright flares. As the flux during the flaring state is one order of magnitude of higher, the total energy release during the flaring state is about 10\% of the energy release in the low state. This may be useful to constrain the underlying physical process for VHE flares.}  

\subsection{Long-term low-state VHE emission}
The LHAASO observations  revealed a long-term VHE emission { with} a flux level of $\sim10^{-12}~\rm erg~cm^{-2}~s^{-1}$.  {The intrinsic spectrum  can be described by a power law, with a possible hardening at 20 TeV.} Although for the episodes of rapid TeV flaring activity an origin in the sub-parsec scale jet and below is preferred, it could well be  that larger scale structures (such as HST-1, the kpc-scale jet and the knot A) contribute to the overall, low-state VHE emission. 
The nuclear environment of M87 does not show evidence of significant sources of soft photons (e.g., from the torus or star formation sites) and therefore the inverse Compton emission is likely dominated by the synchrotron self-Compton (SSC) component from the same population of relativistic electrons that produce the synchrotron  photons. However, a simple one-zone synchrotron and SSC model cannot describe the observed SED easily \citep{2008MNRAS.385L..98T}. 
The reason for that is the following, in order to have the synchrotron component peaking in the infrared (IR) region  and the SSC component close to the TeV band (as required by the relatively flat TeV spectra), one has to assume an unreasonably large Doppler factor, $\delta\gg 10$. The hard TeV spectrum also challenges the one-zone SSC model, since the Klein-Nishina effect leads to a spectral steepening towards  the highest energy end. Indeed, previous detailed  modelings of the low-state emission have found that the SSC model provides a good fit to the SED up to GeV gamma-rays, but it underestimates the VHE emission \citep[e.g.][]{2019A&A...623A...2A,2022ApJ...934..158A}.

To overcome this problem, one should “decouple” the synchrotron and IC components, assuming
that the two components are produced in two different regions, as in the “decelerating jet” model  \citep{2005ApJ...634L..33G} or in the “spine-layer” model  \citep{2008MNRAS.385L..98T}. Other possibility includes  several randomly oriented active regions resulting from reconnection events in the jet \citep[the so-called “jets-in-the-jet” scenario][]{2009MNRAS.395L..29G,2010MNRAS.402.1649G}. An alternative model is a lepto-hadronic emission model \citep{2004A&A...419...89R,2016ApJ...830...81F,2022ApJ...938...79B}, in which three spectral components are involved: a synchrotron-dominated component from radio to X-rays, an inverse Compton-dominated component from X-rays to GeV gamma-rays, and a photohadronic-dominated component in the VHE band.

A spectral hardening seen at 20 TeV (see Figures \ref{SED_WCDA_Obs} and \ref{SED_WCDA}) is statistically insignificant  in the currently available data set.
With  continuous   monitoring by LHAASO for several more years, the source significance will increase continuously and we may be able to test whether the spectral hardening is a statistical fluctuation or a true feature. If the feature is real, it may indicate the presence of an additional component at the high energy,  which could be a  hadronic component. To explain the spectral hardening with hadronic models, the $pp$ process would need a proton spectrum harder than $E^{-2}$, while  the $p\gamma$ process { can more easily}  produce a spectral bump if the target photons peak at a suitable energy. To produce a bump at $\sim20$ TeV via $p\gamma$ process, one would need protons of energy $\sim200$ TeV and the target photons peaking at $\sim1$ keV.

\subsection{Constraining the upper bound of the EBL intensity with the VHE data }

The EBL is not well established with large uncertainties in the flux. Our results allow us to  put a  constraint on the upper bound of the EBL intensity, assuming that the intrinsic spectrum is described by a single power law extending to the maximum energy at 20 TeV.  We adopt a model for the observed spectrum that incorporates a power-law multiplied by the scaled EBL absorption: $dN/dE=N_{0}(E/{\rm 3TeV})^{-\gamma}\times \textrm{exp}(-f_{\rm EBL}\tau(E))$, where $\tau(E)$ is the absorption depth due to EBL for a gamma-ray photon with energy $E$  and  $f_{\rm EBL}$ is the  scaling factor. We use $\tau(E)$ from the \citet{2008A&A...487..837F} as  reference values. Since the absorption for low energy photons below 10 TeV is negligible ($\tau \leq0.1$), the relevant wavelength range\footnote{This wavelength range is obtained from the peak of the pair production cross section at the wavelength for a specific gamma-ray photon of energy E: $\lambda_{\rm max}\simeq1.24(E/{\rm TeV})~\mu m$.} for EBL constraint is $\lambda \ga 10~\mu m$.  
To obtain the upper limit of $f_{\rm EBL}$,  we estimate the likelihood values by varying  the parameter $f_{\rm EBL}$ in fitting  the spectral data observed by LHAASO. The result is shown in Figure \ref{EBL_factor}. We find that the upper limit of $f_{\rm EBL}$ at a 95\% confidence level is about 1.6.
This means that the upper bound of the EBL intensity is a factor of 1.6 larger than the intensity given by  \citet{2008A&A...487..837F} for the wavelength range  $\lambda\ga 10 ~\mu m$.  {We also study the constraint on the \textit{dominguez-upper} model that reflects maximum EBL flux \citep{2011MNRAS.410.2556D}. We find that the upper limit of $f_{\rm EBL}$ at a 95\% confidence level is only  1.09. This means that the upper limit allowed by our TeV data is very close to the maximum EBL flux model. }

 \subsection{Influence of the possible internal absorption on the spectrum}
{  TeV gamma-rays could be absorbed by the radiation fields in the central region of M87.   The observed infrared luminosity of the nucleus of M87 is about  $10^{40}-10^{41} {\rm erg~s^{-1}}$ \citep{2001ApJ...551..206P,2004ApJ...602..116W}. 
Although there are no direct measurements of the size of the infrared source, observations in the microwave band at 43 GHz suggest that its size at millimeter wavelengths is limited by $5\times10^{16}{\rm cm}$, corresponding to 25 $R_s$ for a BH mass of $\sim 6\times10^9 M_\odot$. \citet{2007ApJ...671...85N} calculated the optical depth for gamma-rays produced by the infrared source  assuming that the size of the infrared source is
comparable to that of the microwave source, finding that the nucleus of M87 is effectively transparent for gamma-rays up to an energy of 10 TeV.   \citet{2011ApJ...736...98B} consider  the temperature  distribution of the disk photons of M87 and calculate the  optical depth of gamma-rays for different inclination  angle $\theta$ between the jet and our line of sight. They find that, for $\theta\sim 20^{\circ}$, the optical depth exceeds unity only for photons with energy above 20 TeV.

We perform a re-analysis of the VHE spectra of M87 assuming internal absorption based on the above two models, i.e.,  model 1 for \citet{2007ApJ...671...85N} and model 2 for \citet{2011ApJ...736...98B}. The intrinsic VHE spectra of M87 during the full-time period and the flare period after considering the internal absorption are shown in Figure \ref{InternalAbs}. It is found that the  spectra become harder with a spectral index  being $1.86\pm0.21$ for model 1 and $2.06\pm0.20$ for model 2 during the full-time period. During the flare period, the spectral index  is $2.26\pm0.31$ for model 1 and $2.41\pm0.30$ for model 2. The excess at 20 TeV becomes more pronounced in the VHE spectra  during the full-time period. Note that, in the calculation, we have assumed that the emitting region  of the VHE emission during the low state is also close the AGN core, similar to the case during the flare state.}

\section{Summary}
\label{sect:sum}
We conducted  three-year continuous observations of M87 from 2021 to 2024 with LHAASO, which yield a detection of a signal with a statistical significance of $\sim9\sigma$.  The LHAASO observations  revealed a long-term VHE emission  with a flux level of $\sim10^{-12}~\rm erg~cm^{-2}~s^{-1}$. The observed energy spectrum extends to 20 TeV, showing a  cutoff shape at higher energies. Assuming that the intrinsic spectrum is described by a single power law up to 20 TeV,  we obtained a stringent upper bound on the EBL intensity.  

No clear variability is observed in the long-term light curve on three-month time bins of the LHAASO data. A strong flare,  lasting {eight} days, is found in early 2022 on a two-day time bin analysis. The flux during the flare is higher than the long-term average flux by one order of magnitude.  The light curve of the flare is described by a two-sided exponential function with a rise time of $\tau_{d}^{\rm rise} = 1.05\pm0.49$~days  and a decay time of $\tau_{d}^{\rm decay} = 2.17\pm0.58$~days. Unlike the previous flares, the 2022 flare rises faster and decays slower. The rapid variability suggest that the VHE emission arises from a compact emission region with a size of $R\sim 2.7\times10^{15}\delta\, {\rm cm}$ ($\delta  {\la 3}$), corresponding to only a few Schwarzschild radii of the central supermassive black hole.  This suggests that the relevant non-thermal particle acceleration and emission processes could take place in the vicinity of the central black hole. We also found a possible  GeV flare in the Fermi-LAT data during the TeV flare period. 


The continuously monitoring of M87 with LHAASO makes it possible, for the first time, to give an unbiased estimate of the duty cycle of bright VHE flares. One strong flare observed during the three-year observations  indicates  a duty cycle of $\sim 1\%$ for VHE flares with a flux reaching $ 10^{-11}{\rm~erg~cm^{-2}~s^{-1}}$.

\begin{acknowledgments}

We would like to thank all staff members who work at the LHAASO site above 4400 m a.s.l. year round to maintain the detector and keep the water recycling system, electricity power supply and other components of the experiment operating smoothly. We are grateful to Chengdu Management Committee of Tianfu New Area for the constant financial support for research with LHAASO data. 
The work is supported by the NSFC under grants Nos. 12333006, 12203022, 12121003, and the Natural Science Foundation of Jiangsu Province grant BK20220757, and in Thailand by the National Science and Technology Development Agency
(NSTDA) and the National Research Council of Thailand (NRCT) under the High-Potential Research Team Grant Program (N42A650868).

\end{acknowledgments}



\bibliography{reference}{}
\bibliographystyle{aasjournal}

\begin{figure*}
\centering
\includegraphics[angle=0,scale=0.28]{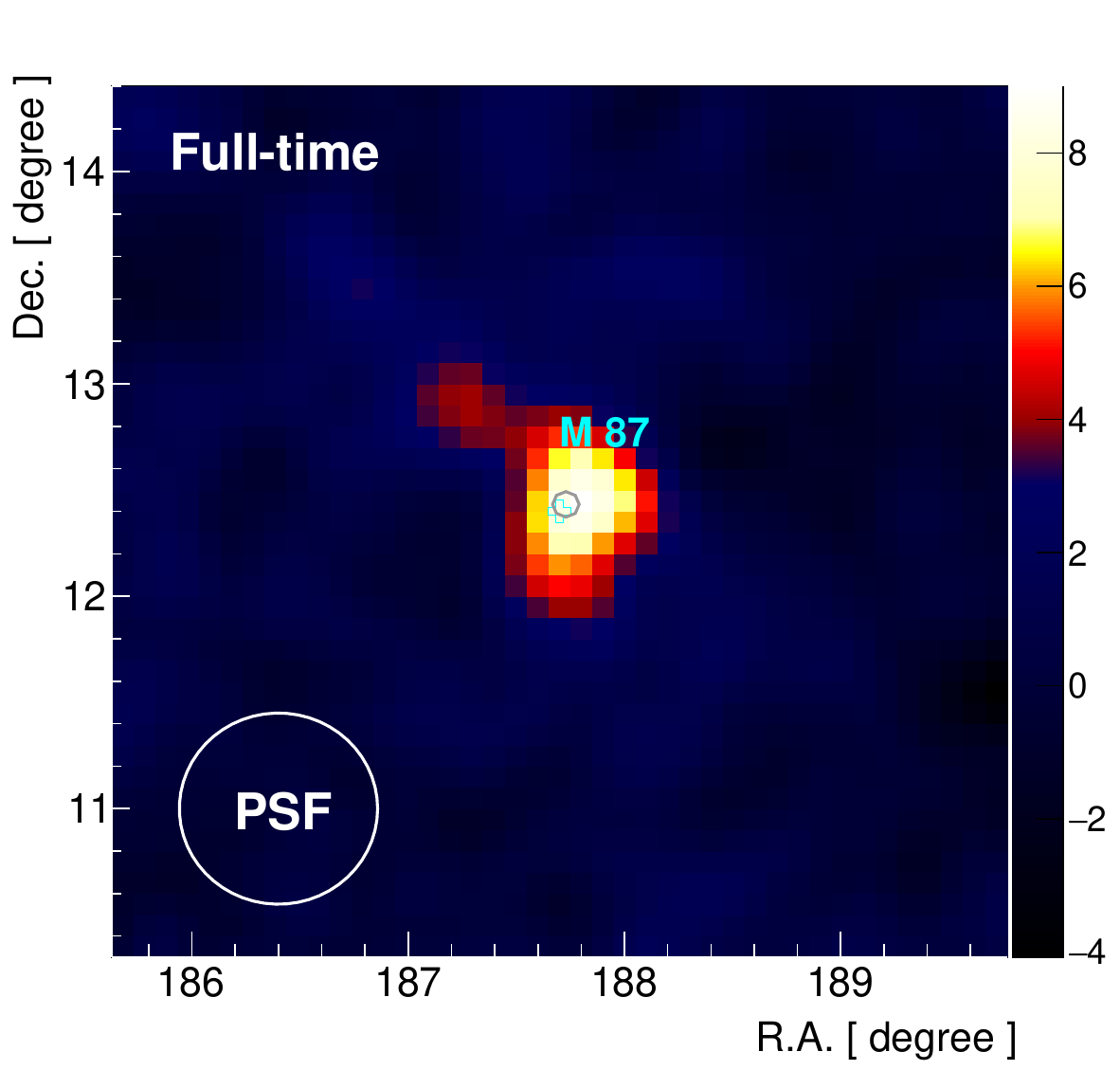}
\includegraphics[angle=0,scale=0.28]{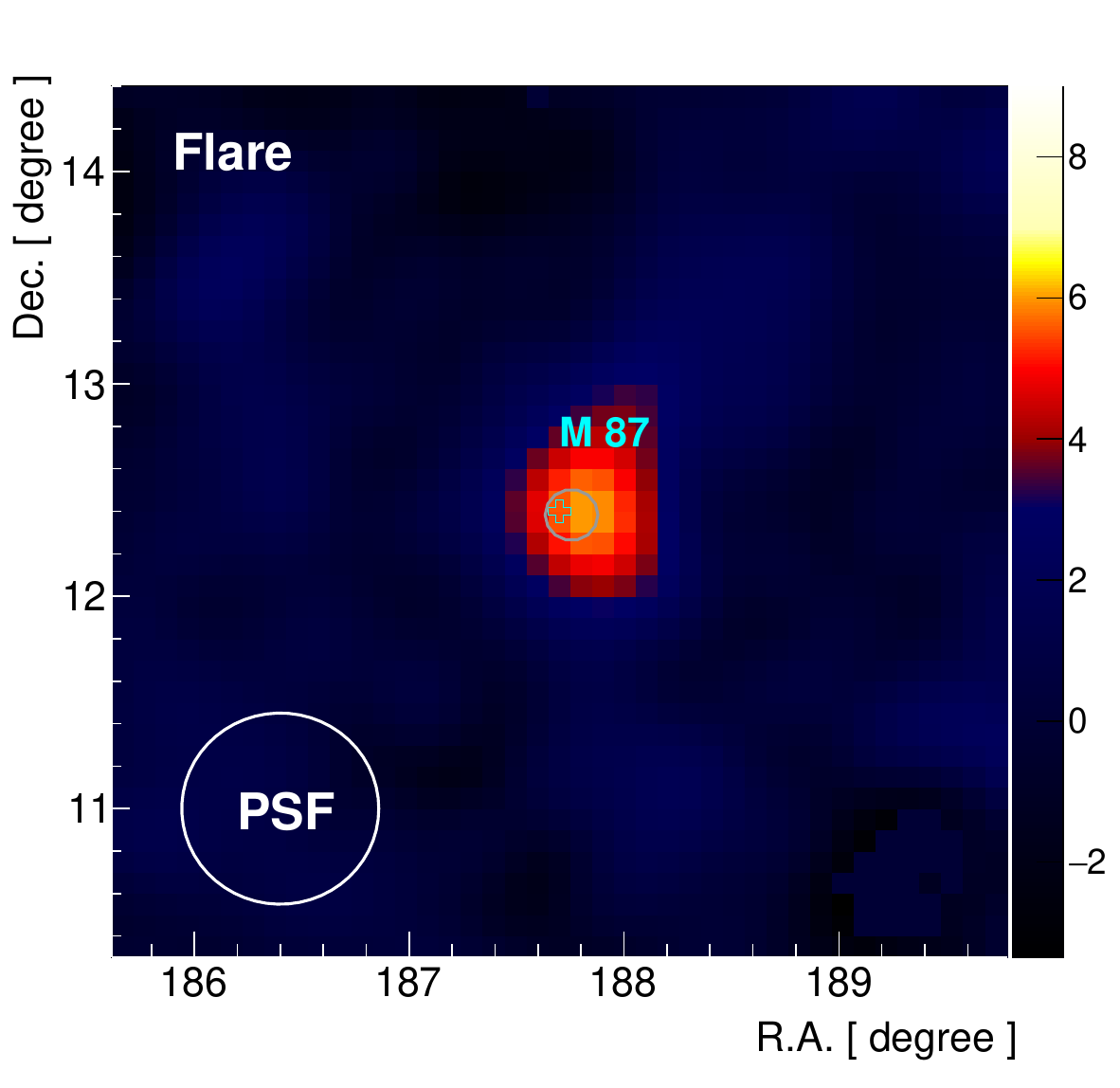}
\includegraphics[angle=0,scale=0.28]{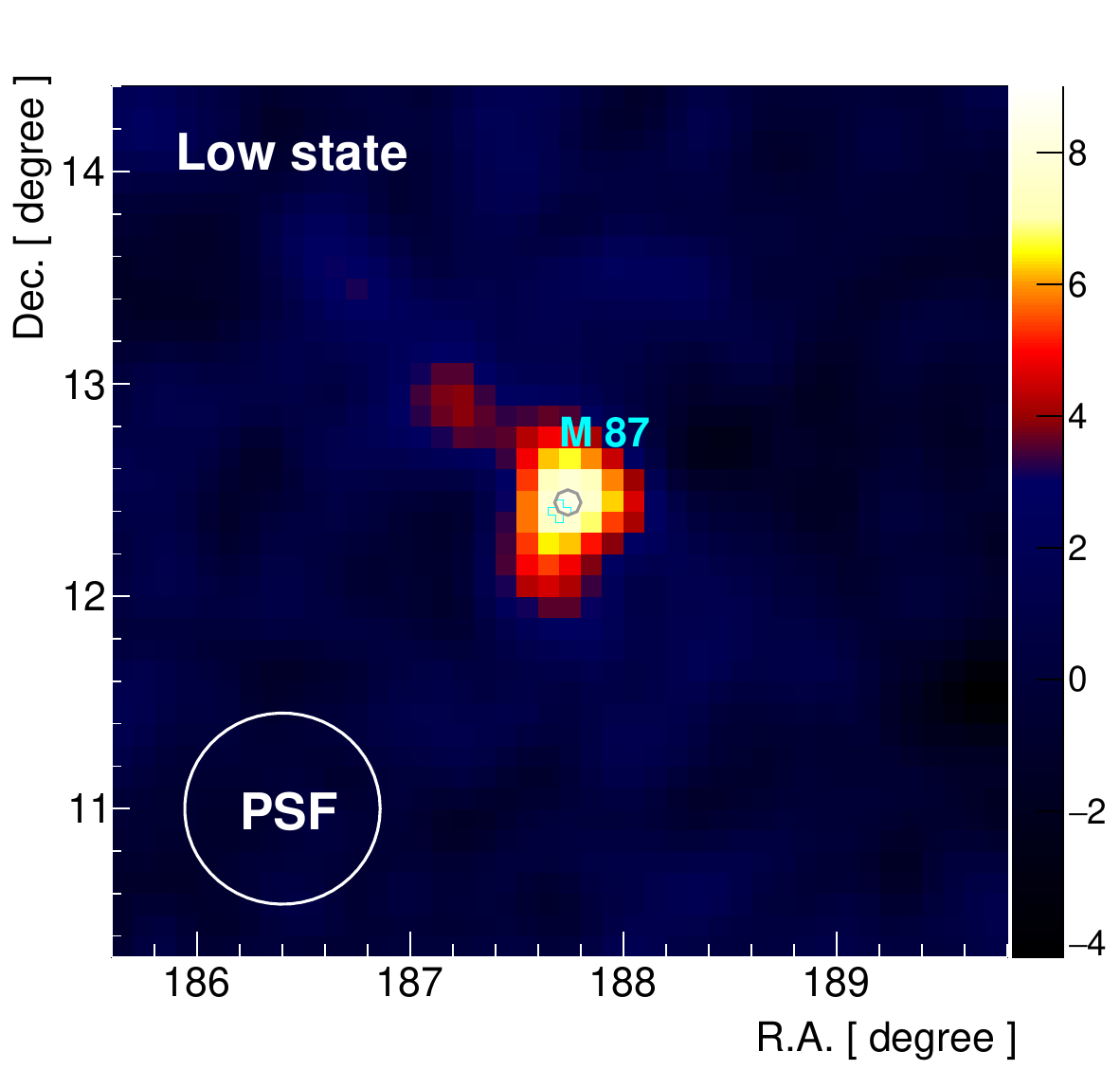}
\caption{Significance maps of $2^{\circ}\times2^{\circ}$ region around M87 by { LHAASO-WCDA} for full-time period (left panel), flare state (middle panel) and low state (right panel), respectively. M87 is marked by cyan open cross. The gray circle indicates the 95\% position error of the source. The white circle at the bottom-left corner shows the size of the LHAASO PSF (68\% containment).}
\label{SigMap_WCDA}
\end{figure*}

\begin{figure*}
\centering
\includegraphics[angle=0,scale=0.40]{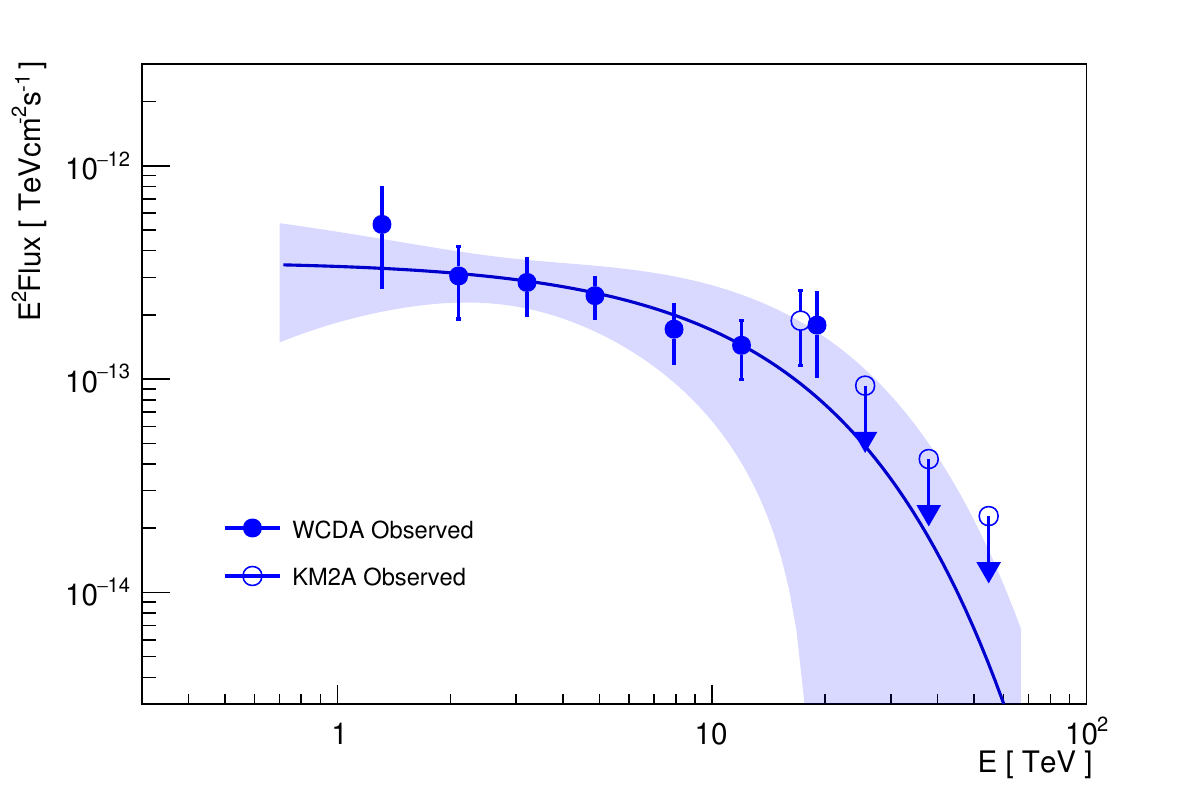}
\includegraphics[angle=0,scale=0.40]{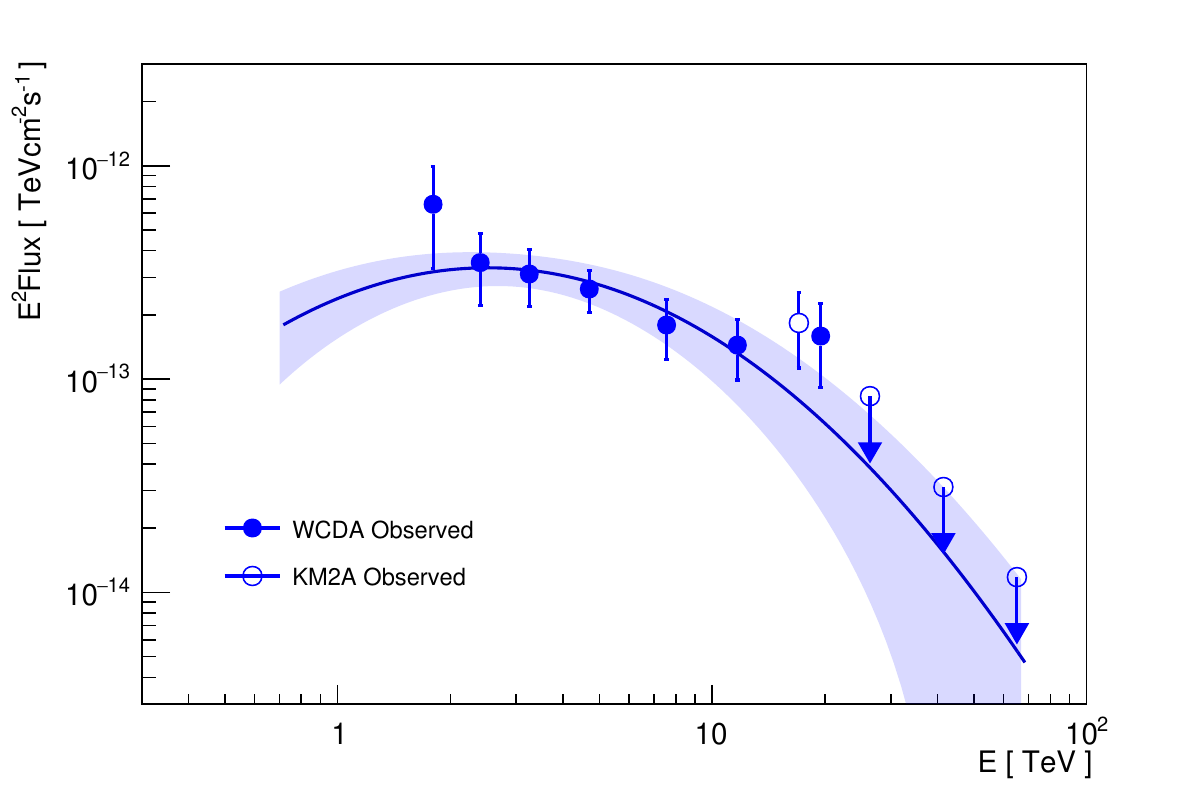}
\caption{Observed VHE spectra of M87  by LHAASO during the full-time period assuming the PLEC function and  LP function, respectively. The solid lines indicate the best-fitting results and the blue shaded regions indicate the $1\sigma$ statistical error. Left: The PLEC function is adopted to fit the observational data. Right: The LP function is adopted to fit the observational data. }
\label{SED_WCDA_Obs}
\end{figure*}

\begin{figure*}
\centering
\includegraphics[angle=0,scale=0.28]{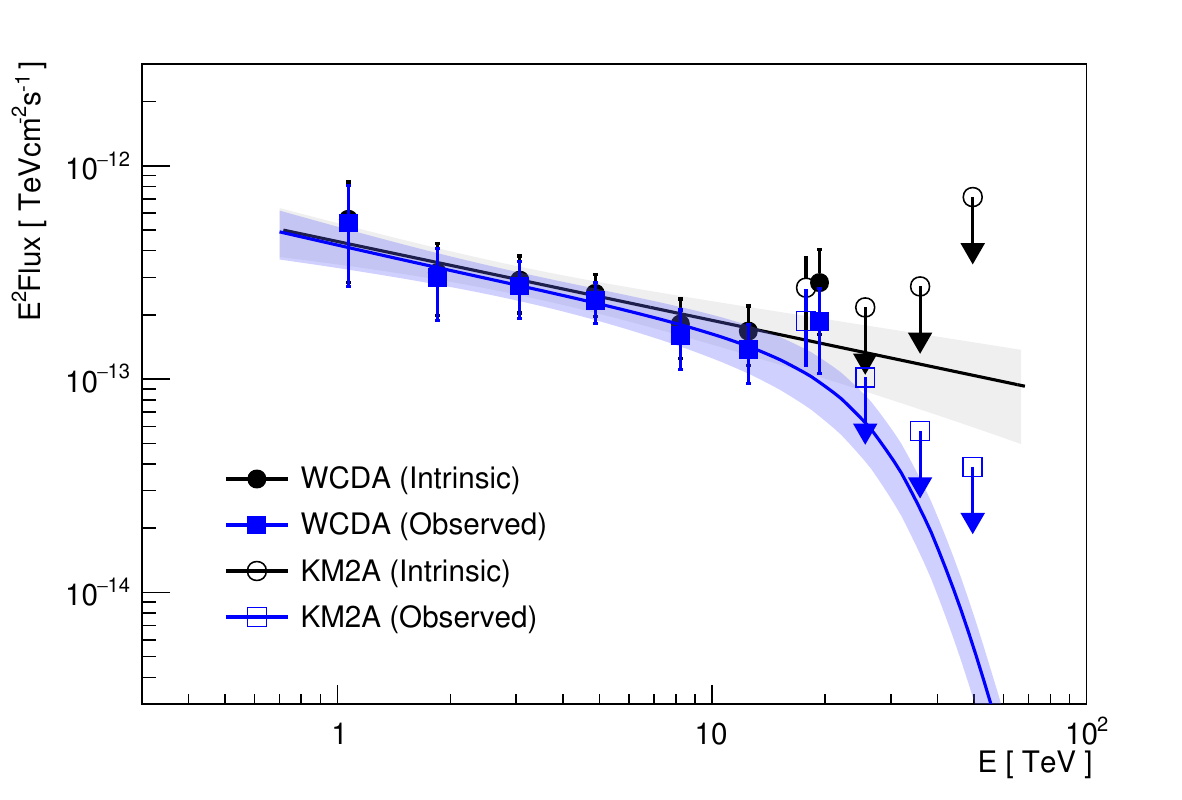}
\includegraphics[angle=0,scale=0.28]{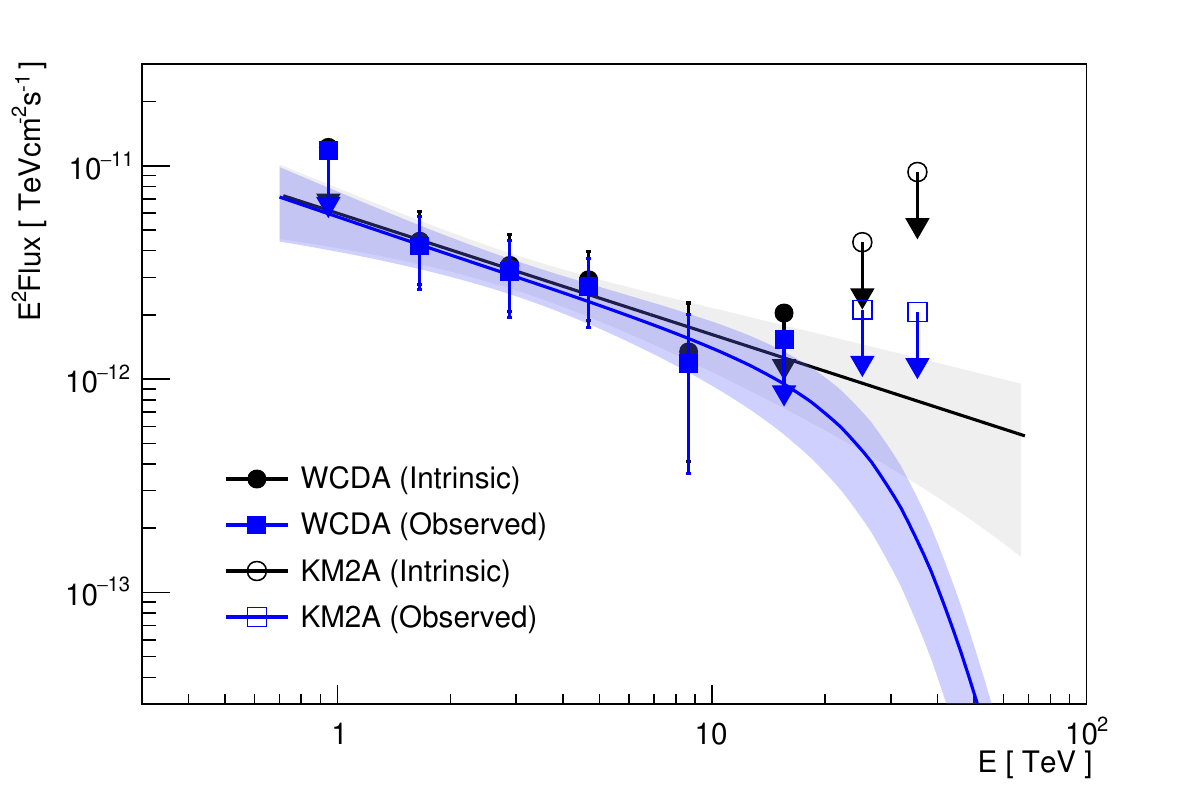}
\includegraphics[angle=0,scale=0.28]{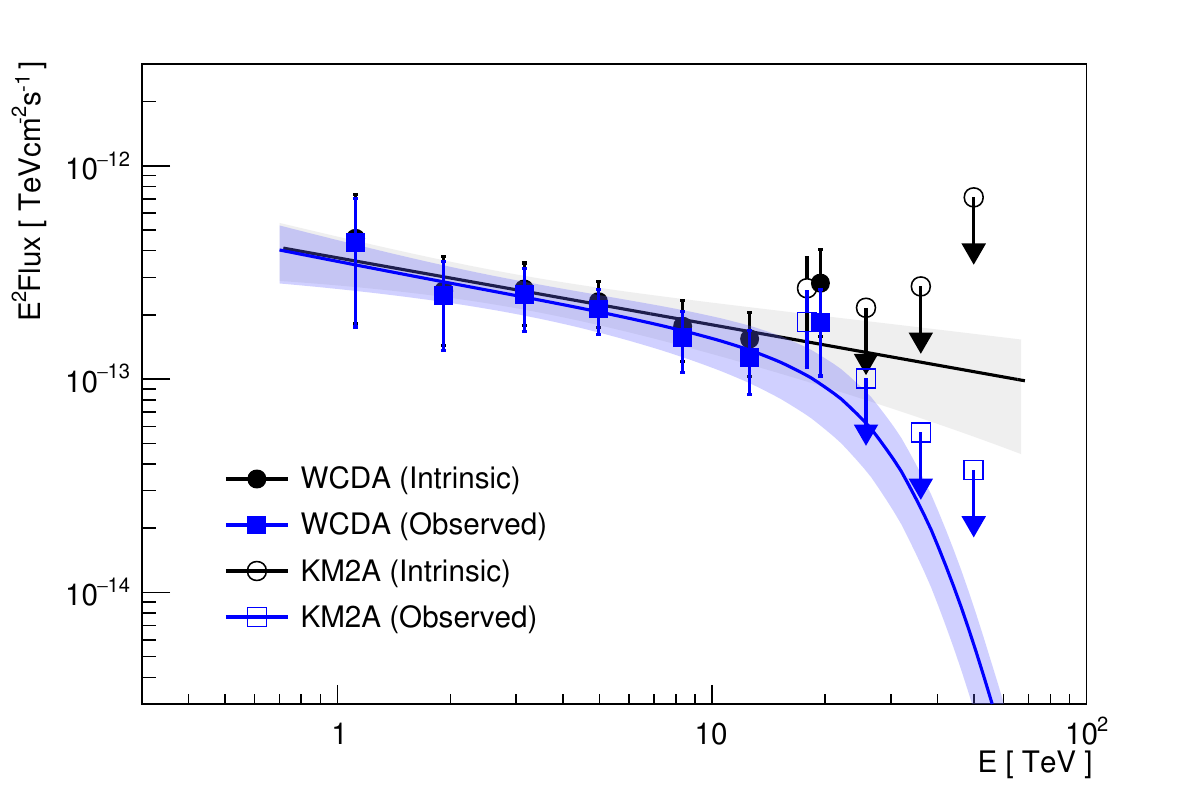}
\caption{VHE spectra of M87  during the full-time period (left panel), the flare state (middle panel) and the low state (right panel), assuming a single power-law model for the intrinsic spectra and a EBL absorption model of \citet{2008A&A...487..837F}.
The gray and blue shaded regions indicate $1\sigma$ statistical error for the intrinsic spectra and PLxEBL spectra, respectively.}
\label{SED_WCDA}
\end{figure*}

\begin{figure*}
\begin{minipage}{0.45\textwidth}
\centering
\includegraphics[angle=0,scale=0.46]{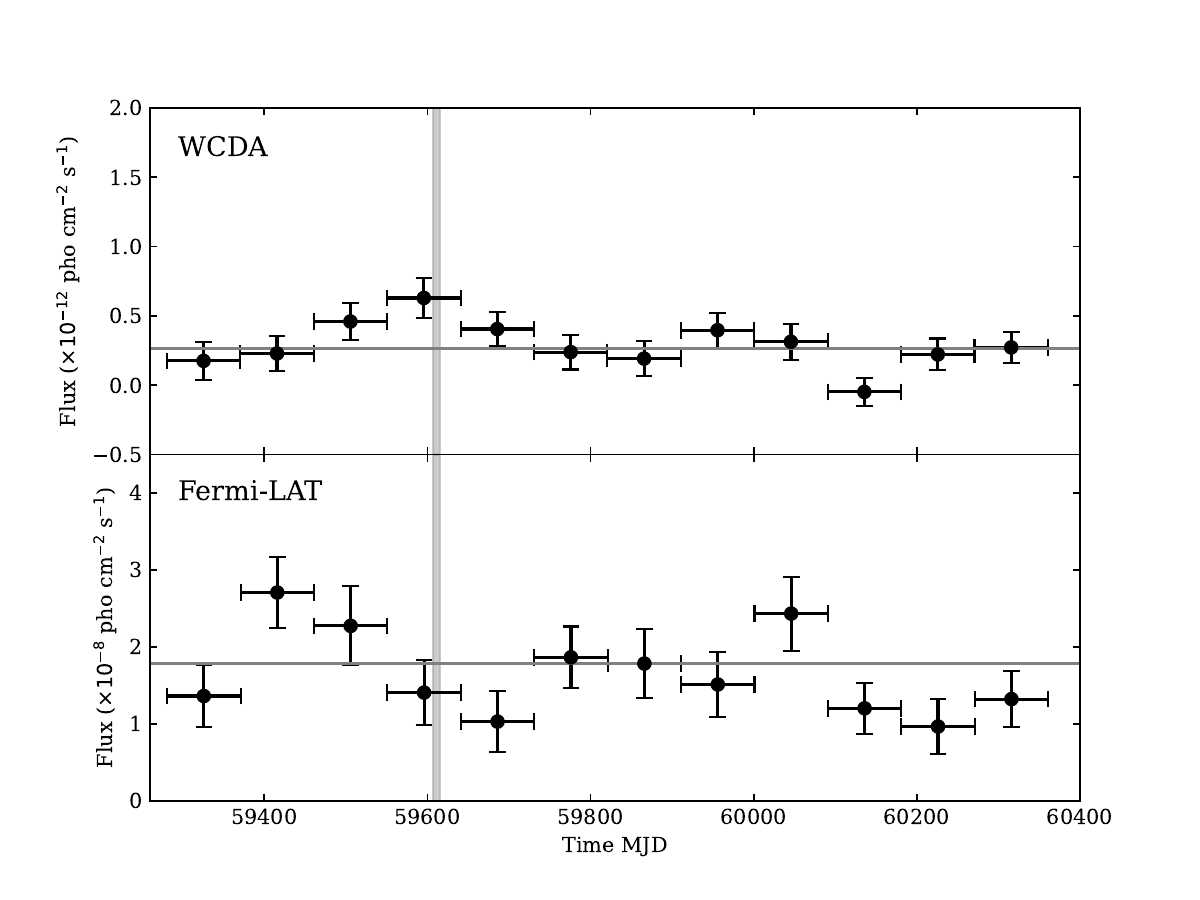}
\end{minipage}
\hspace{0.5cm} 
\begin{minipage}{0.45\textwidth}
\centering
\includegraphics[angle=0,scale=0.46]{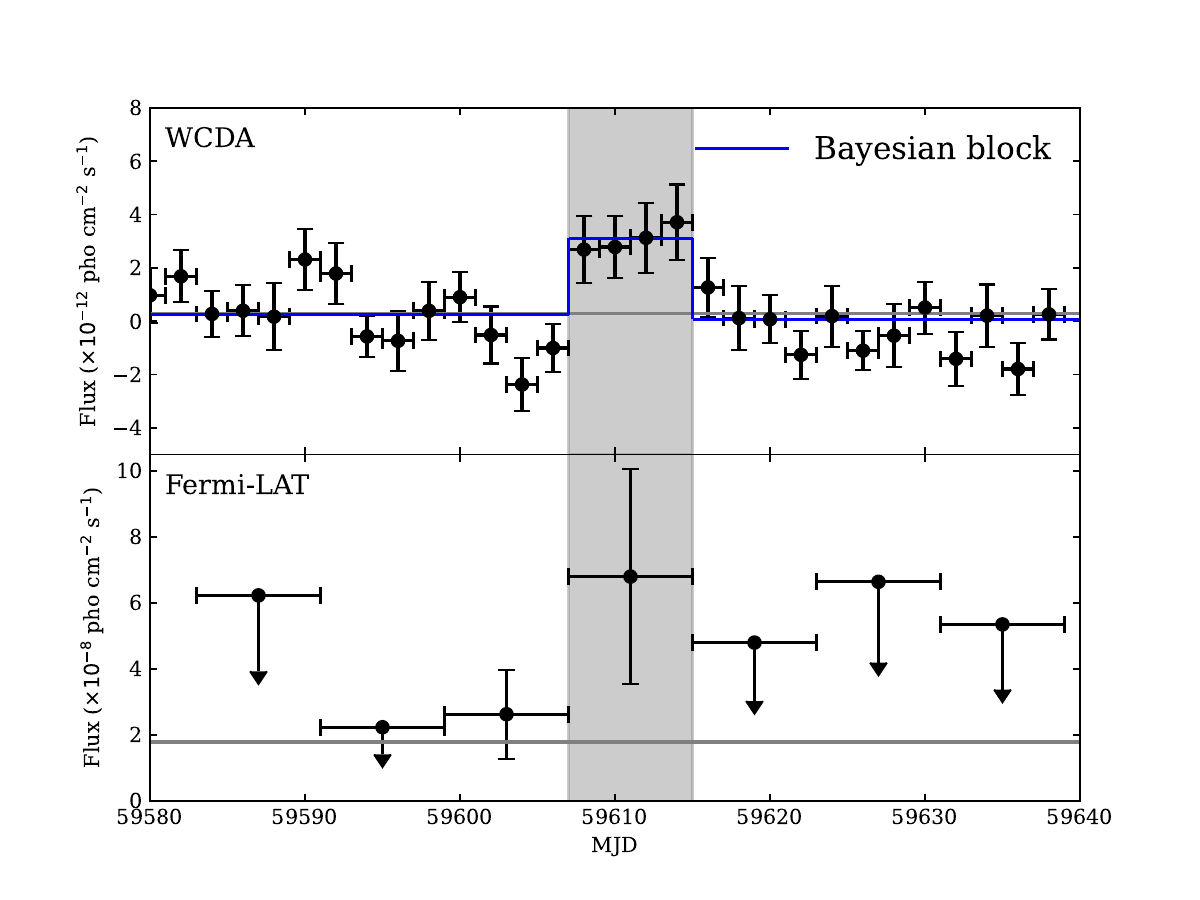}
\end{minipage}
\caption{Light curves of LHAASO-WCDA (top panels) and Fermi-LAT (bottom panels). 
Left panel: long-term light curves (three-month bin) of LHAASO-WCDA and Fermi-LAT for M87.
Right panel: zoom-in light curves around the flare period. The light curve of LHAASO-WCDA is obtained with two-day bin and that of Fermi-LAT is obtained with 8-day bin. $1\sigma$ upper limit is reported when $\rm TS<9$ for the LAT light curve data. 
The LHAASO-WCDA flare period is shown as the gray vertical shaded region.
The blue line in the top panel indicates the bayesian block and the gray lines show the long-term  average flux.
}
\label{LC1}
\end{figure*}

\begin{figure*}
\centering
\includegraphics[angle=0,scale=0.6]{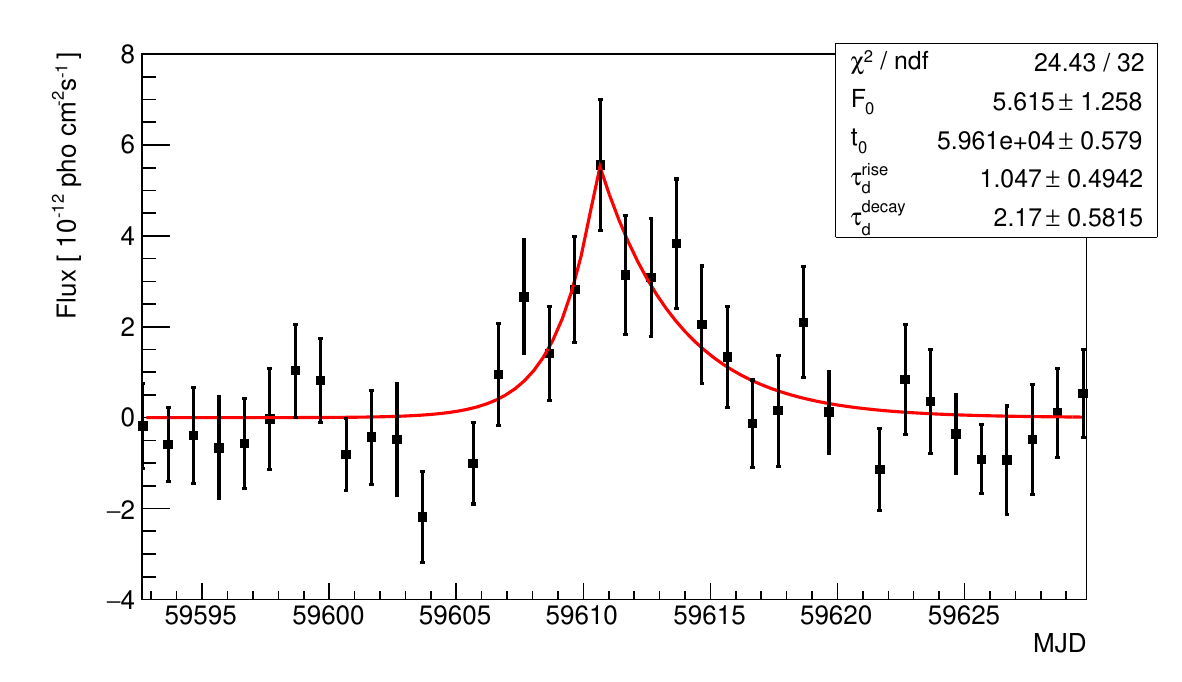}
\caption{Fitting of the one-day bin light curve (range from MJD 59593 to MJD 59629) of the  VHE flare measured by LHAASO-WCDA with a two-sided exponential function (the red line). Fitting parameters are $F_0=(5.62\pm1.26)\times10^{-12}~ \rm TeV^{-1}~cm^{-2}~s^{-1}$, $t_0= \rm MJD~59610\pm0.58$, $\tau_{d}^{\rm rise}=1.05\pm0.49$~days and $\tau_{d}^{\rm decay}=2.17\pm0.58$~days.
}
\label{LC2}
\end{figure*}

\begin{figure*}
\centering
\includegraphics[angle=0,scale=0.7]{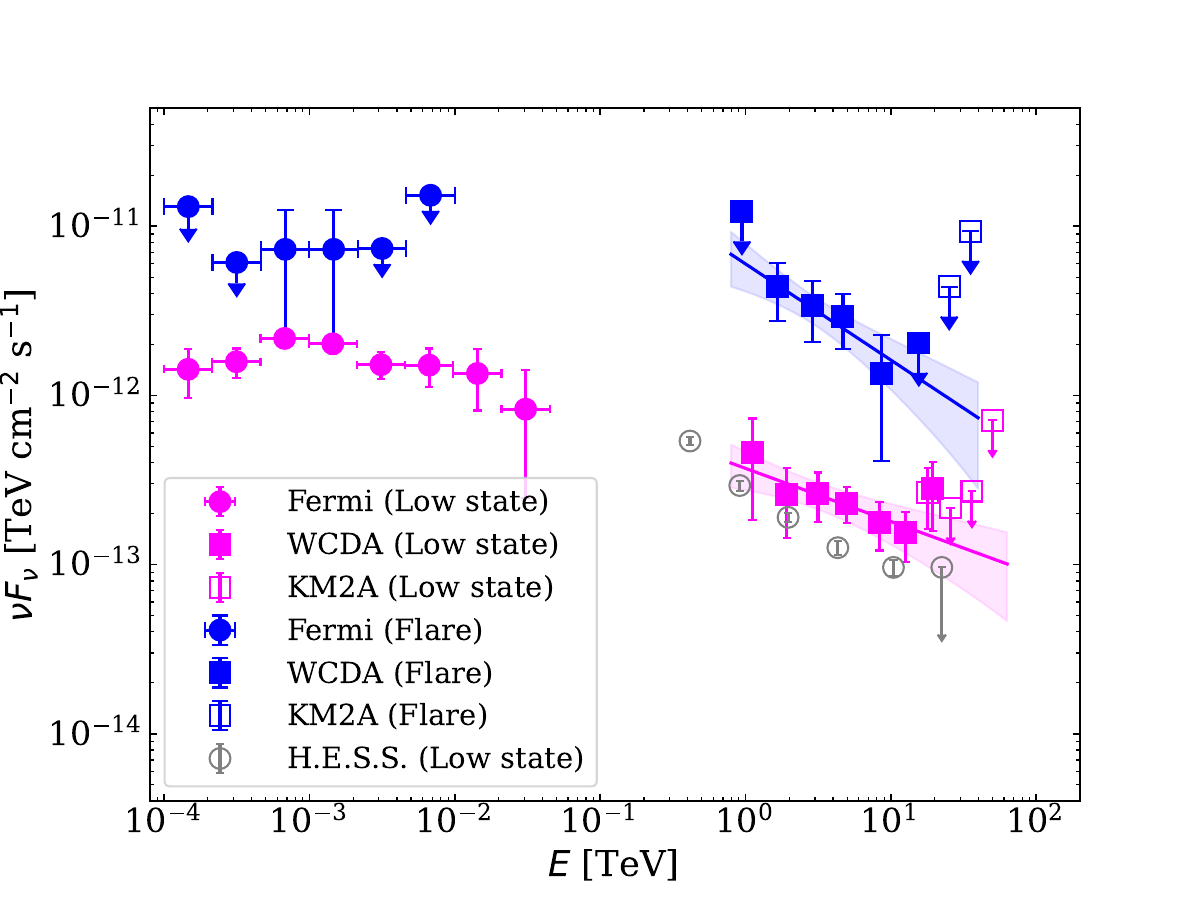}
\caption{Gamma-ray SED of M87 for the low state (magenta data points) and flare state (blue data points), respectively. The flare state encompasses data from MJD 59607 to MJD 59615. 
{  The VHE data of LHAASO show the intrinsic spectra, and the shaded regions indicate $1\sigma$ statistical error for the fitting of the intrinsic spectra.}
The gray data points represent the observation of H.E.S.S. in the low state \citep{2023A&A...675A.138H}.}
\label{SED_0}
\end{figure*}

\begin{table}[!htp]
\caption{Best-fit parameters of the intrinsic spectrum for the full-time LHAASO data based on  three EBL models.}
\label{tab1:EBL_SED}
\begin{center}
\begin{tabular}{lccccc}
\hline
\hline
 EBL model &  $N_0 @{\rm 3TeV}$  &  $\gamma$   &  $E_{\gamma,\rm{max}}$   &  TS  &  $\chi^{2}/{\rm ndf}$  \\
          & ($\rm{10^{-13}~TeV^{-1}~cm^{-2}~s^{-1}}$) &  & (TeV) &  & \\
\hline
\textit{dominguez-upper} & $0.323\pm0.055$ & $2.25\pm0.18$ & 18.4 & 87.5 & 3.47/4\\
\textit{franceschini} & $0.327\pm0.050$ & $2.37\pm0.14$ & 19.4 & 87.4 & 3.00/4\\
\textit{kneiske} & $0.330\pm0.044$ & $2.56\pm0.12$ & 20.6 & 84.4 & 3.97/4\\
\hline
\hline
\end{tabular}\\
\end{center}
\end{table}

\begin{table}[!htp]
\caption{Best-fit spectral parameters of the LHAASO data for different time intervals.}
\label{tab1:OverallSED}
\begin{center}
\begin{tabular}{lccccc}
\hline
\hline
Time interval & State  & TS &  $N_0 @{\rm 3TeV}$  &  $\gamma$ & Flux  \\
 (MJD)  &    &  & ($\rm{TeV^{-1}~cm^{-2}~s^{-1}}$) &  & ($\rm{pho~cm^{-2}~s^{-1}}$)\\
\hline
59281--60385 & Full-time & 87.4 &(3.27$\pm$0.50)$\times10^{-14}$ & 2.37$\pm$0.14 &  (3.17$\pm$0.49)$\times10^{-13}$ \\
59607--59615 & Flare & 40.1 &(3.56$\pm$0.67)$\times10^{-13}$ & 2.57$\pm$0.23 &  (3.78$\pm$0.71)$\times10^{-12}$ \\
59281--59606 & \multirow{2}{*}{Low} & \multirow{2}{*}{76.0} &\multirow{2}{*}{(2.91$\pm$0.53)$\times10^{-14}$} & \multirow{2}{*}{2.31$\pm$0.17} &  \multirow{2}{*}{(2.76$\pm$0.50)$\times10^{-13}$} \\
59616--60385 & & & & & \\
\hline
\hline
\end{tabular}\\
\end{center}
\end{table}


\begin{table}[!htp]
\caption{Best-fit spectral parameters of the Fermi-LAT data.}
\label{tab2:latspec}
\begin{center}
\begin{tabular}{lcccc}
\hline
\hline
Time interval  & TS &  $N_0$  &  $\Gamma_{LAT}$ & Flux  \\
 (MJD)  &    & ($\rm{ph~cm^{-2}~s^{-1}}$) &  & ($\rm{pho~cm^{-2}~s^{-1}}$)\\
\hline
54682.65--60385  & 2434.31 & (1.67$\pm$0.05)$\times10^{-12}$ & 2.05$\pm$0.03 &  (1.79$\pm$0.10)$\times10^{-8}$ \\
59281--60385  & 505.29 & (1.78$\pm$0.13)$\times10^{-12}$ & 2.06$\pm$0.05 &  (1.93$\pm$0.22)$\times10^{-8}$ \\
59607--59615  & 9.50 & (2.75$\pm$1.48)$\times10^{-12}$ & 2.40$\pm$0.35 &  (6.80$\pm$3.26)$\times10^{-8}$ \\
\hline
\hline
\end{tabular}\\
\end{center}
\end{table}

\begin{figure}
\centering
\includegraphics[angle=0,scale=0.40]{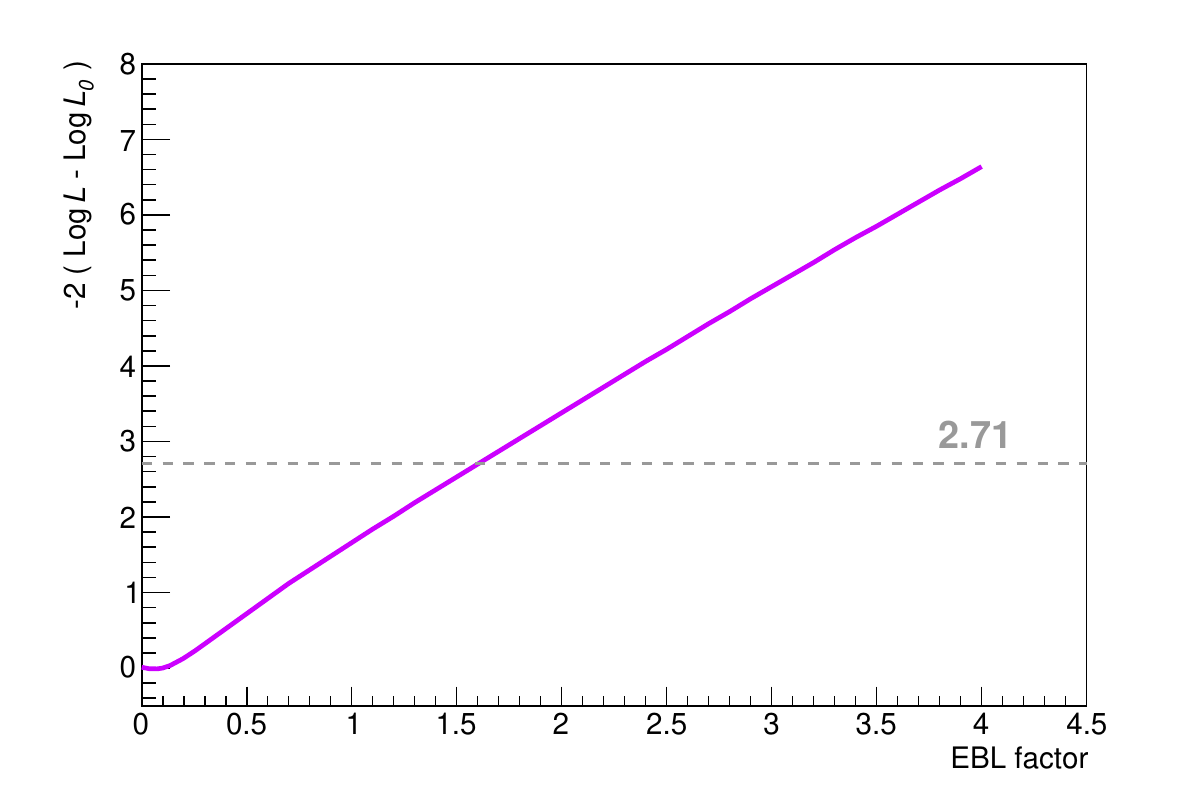}
\includegraphics[angle=0,scale=0.40]{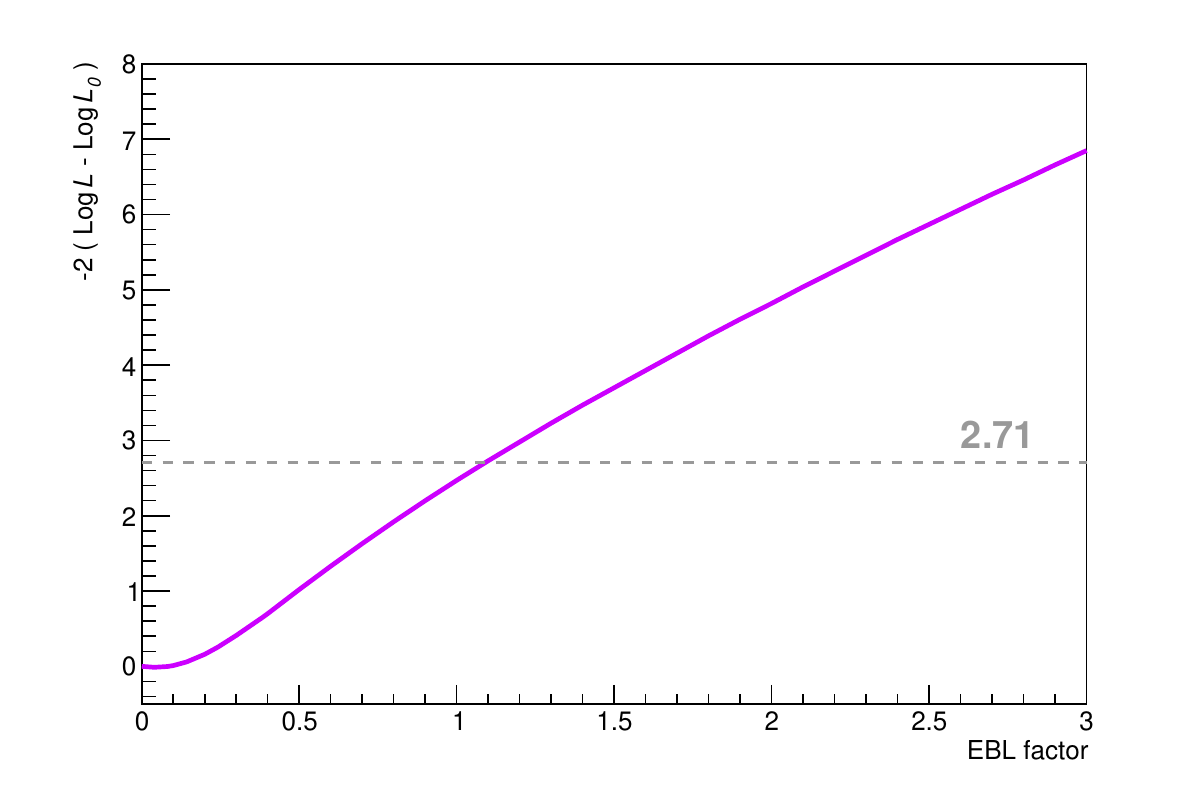}
\caption{The likelihood profile on fitting the spectral data
observed by LHAASO for a
varying  scale factor $f_{\rm EBL}$  based on the EBL model of \citet{2008A&A...487..837F}  (left panel)  and the EBL model of \citet{2011MNRAS.410.2556D} (right panel). The grey dashed line represents $-2({\rm log}L-{\rm log}L_{0})=2.71$,  which corresponds to a 95\% confidence level (where $L$ is the likelihood for different scale factors $f_{\rm EBL}$ and $L_{0}$ is the maximum likelihood ).} 
\label{EBL_factor}
\end{figure}

\begin{figure}
\centering
\includegraphics[angle=0,scale=0.45]{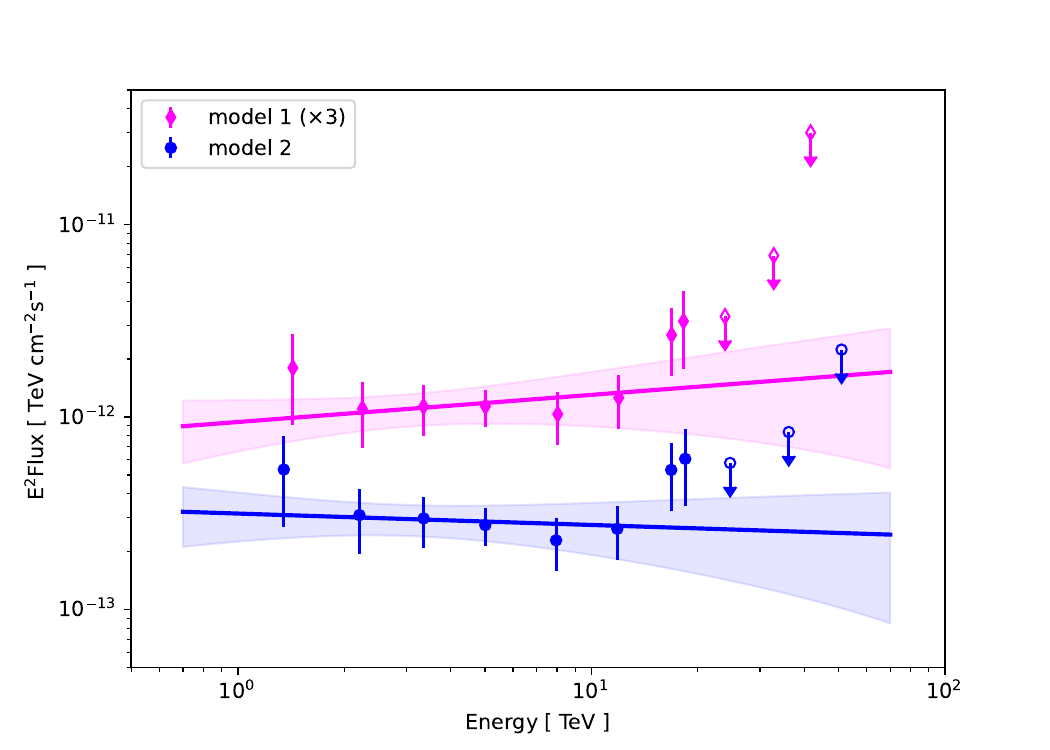}
\includegraphics[angle=0,scale=0.45]{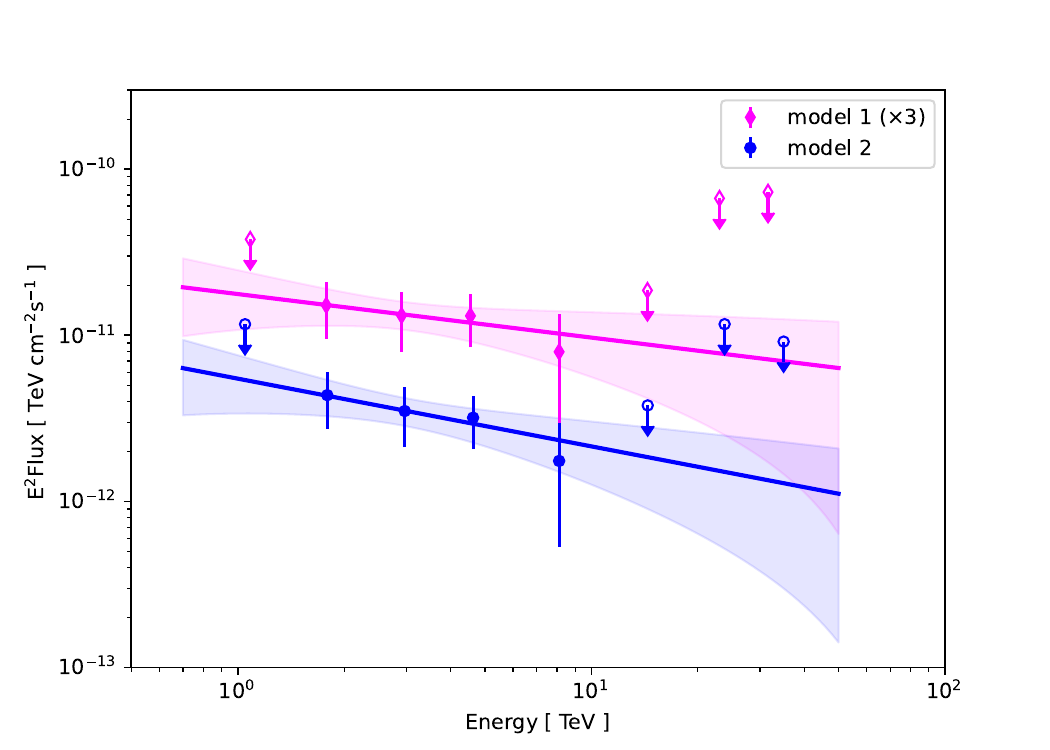}
\caption{Intrinsic VHE spectra of M87 during the full-time period (left panel) and the flare period (right panel) after considering the internal absorption modeled by \citet{2007ApJ...671...85N} (model 1) and \citet{Brodatzki_2011} (model 2).} 
\label{InternalAbs}
\end{figure}

\appendix
\section{LHAASO Data Reduction}
\label{sect:app_haaso}
The dataset utilized in this analysis was mainly sourced from the WCDA at LHAASO. For this study, we restricted our analysis to the events with a zenith angle of less than 50 degrees to ensure high-quality reconstruction data. The selected events are subsequently categorized into seven groups based on the effective number of triggered PMT units, denoted as $N_{\rm hit}$, with ranges [60-100), [100-200), [200-300), [300-500), [500-700), [700-1000) and [1000, 2000]. Additionally, a gamma/proton separation parameter $\mathcal{P}_{\rm{ciness}}$ is employed to effectively identify gamma-like events. $\mathcal{P}_{\rm{ciness}}$ cut of seven $N_{\textrm{hit}}$ segments are 1.02, 0.90, 0.88, 0.88, 0.84, 0.84 and 0.84. Ultimately, the effective live-time for observations of M87 was recorded as 1026 days, with an approximate count of $9.61\times10^{9}$ gamma-like events.
The event maps are created as histograms of the arrival direction of the reconstructed events, after converting the arrival direction of the shower from local coordinates to equatorial coordinates. The background maps are computed using a direct integration method \citep{Fleysher_2004}. For this analysis, a sliding time window of 10 hours was used to estimate the detector acceptance, with an integration time set to 4 hour to estimate the relative back-ground. Events within the region of the Galactic plane and LHAASO 1st catalog gamma-ray sources (with a spatial angle less than 5 degrees) were excluded from the background estimation. The excess map is then obtained by subtracting the background map from the event data map.

For the spectral fitting, we assume a simple power-law function for the intrinsic spectrum. The photon attenuation derived from the three EBL models (e.g., the \textit{kneiske} model \citep{2010A&A...515A..19K}, \textit{franceschini} model \citep{2008A&A...487..837F}, and \textit{dominguez-upper} model \citep{2011MNRAS.410.2556D}) as a function of energy at the distance z = 0.0042 is applied to correct the detector effective area, via re-weighting every simulated event with the survival probability $\mathcal{P}(E)=e^{-\tau(E)}$ from the EBL attenuation, where $\tau(E)$ is the optical depth due to the EBL absorption at the photon energy $E$. The results are summarized in Table \ref{tab1:EBL_SED}, indicating that the impact of different EBL models on the maximum energy is not significant.

\section{Fermi-LAT Data Reduction}
\label{sect:app_lat}

The Pass 8 data of the M87 region, covering 15.5 yr (MJD 54683--60385) are used for our analysis. We select the region of interest (ROI) centered at the radio position of M87 with a radius of 15$\degr$. The publicly available software \textit{Fermipy} (version v1.1)\citep{Wood2017} and \textit{Fermitools} \citep[version 2.2.0,][]{2019ascl.soft05011F} are used to perform the data analysis with the binned maximum-likelihood method. The $\gamma$-ray events in the energy range of 0.1--100 GeV are selected with a standard data quality selection criteria of “(DATA\_QUAL\textgreater0)\&\&(LAT\_CONFIG==1)”. We binned the data with a pixel size of $0.1\degr$ and twelve energy bins per decade. To reduce the $\gamma$-ray contamination from the Earth limb, a maximum zenith angle of 90$\degr$ is set. The event class $P8R3\_SOURCE$ (“evclass=128”), and corresponding instrument response functions (IRFs) P8R3\_SOURCE\_V3 is used in our analysis. 
For background model, we include the diffuse Galactic interstellar emission (IEM, $gll\_iem\_v07.fits$),  isotropic emission (``$iso\_P8R3\_SOURCE\_V3\_v1.txt$'' ) and all sources listed in the fourth \textit{Fermi}-LAT catalog \citep{abdollahi2020fermi}.
The parameters of all the sources within $5\degr$ of the center, IEM and isotropic emission are set free. The spectrum of M87 is modelled with a power-law function i.e., $dN(E)dE=N_0(E/E_0)^{-\Gamma_{LAT}}$, where $\Gamma_{LAT}$ is the photon index.
The light curves with different intervals are obtained by { freeing} the normalisation of M87 and { fixing} all the other spectral parameters to the best-fit values obtained from 15.5-yr analysis.

\begin{table}[!htp]
\caption{Fitting results of the intrinsic spectrum with different models.}
\label{tab4:SED_model}
\begin{center}
\begin{tabular}{lcccccc}
\hline
\hline
SED model &  $N_0 @{\rm 3TeV}$  &  $\gamma$ & $\beta$  ($E_{\rm cut}/{\rm TeV})$ & $E_{\gamma, \rm{max}}$ & $\rm {\Delta TS}$ \\
& ($\rm{10^{-13}~TeV^{-1}~cm^{-2}~s^{-1}}$) & & & (TeV) &\\
\hline 
PL & $0.327\pm0.050$ & $2.37\pm0.14$ & -- &19.4 & 0\\
LP & $0.327\pm0.056$ & $2.37\pm0.17$ & $\beta=0.00\pm0.78$ & 19.4 & -0.12 \\
PLEC & $0.337\pm0.058$ & $2.30\pm0.18$ & $E_{\rm cut}=117\pm309$ & 19.1 & 0.07 \\
\hline
\hline
\end{tabular}\\
\end{center}
\end{table}

\end{document}